\documentclass{jfm}

\usepackage{soul}
\usepackage{graphicx}
\usepackage{newtxtext}
\usepackage{newtxmath}
\usepackage{natbib}
\usepackage{hyperref}
\usepackage{float}
\usepackage{placeins}
\usepackage{caption}
\usepackage{upgreek}

\usepackage[normalem]{ulem}
\hypersetup{
    colorlinks = true,
    urlcolor   = blue,
    citecolor  = black,
}

\newcommand{\RomanNumeralCaps}[1]

\title{Dynamic equilibrium of electrochemical bubbles growing on micro-electrodes}
\author{Mengyuan Huang\aff{1,2}, 
 Chao Sun\aff{3},    
 Kerstin Eckert\aff{1,2},
 Xianren Zhang\aff{4}
 \corresp{\email{zhangxr@buct.edu.cn}}
 \and Gerd Mutschke\aff{2}
 \corresp{\email{g.mutschke@hzdr.de}}
}

\affiliation{
\aff{1}Institute of Process Engineering and Environmental Technology, Technical University of Dresden, 01069 Dresden, Germany 
\aff{2}Institute of Fluid Dynamics, Helmholtz-Zentrum Dresden-Rossendorf (HZDR), 01328 Dresden, Germany
\aff{3}New Cornerstone Science Laboratory, Center for Combustion Energy,
Key Laboratory for Thermal Science and Power Engineering of Ministry of Education, Department of Energy and Power Engineering, Tsinghua University, 100084 Beijing, China
\aff{4}State Key Laboratory of Organic-Inorganic Composites, Beijing University of Chemical Technology, 100029 Beijing, China
}

\begin{document}
\maketitle

\begin{abstract}
In gas evolving electrolysis, bubbles grow at electrodes due to a diffusive influx from oversaturation generated locally in the electrolyte by the electrode reaction.
When considering electrodes of micrometer-size resembling catalytic islands, bubbles are found to approach dynamic equilibrium states at 
which they neither grow nor shrink.
Such equilibrium states are found at low oversaturation for both, pinning and expanding wetting regimes of the bubbles
and are based on the balance of local influx near the bubble foot and global outflux.
Unlike the stability of pinned nano-bubbles studied earlier, 
the Laplace pressure plays a minor role only.
We extend the analytical solution of \mbox{\cite{zhang_lohse_2023}} by taking into account the non-uniform distribution of dissolved gas around the bubble obtained from direct numerical simulation.
This allows us to identify the parameter regions of 
bubble growth, dissolution and dynamic equilibrium as well as to analyze the stability of the equilibrium states.  Finally, we draw conclusions on how to possibly enhance the efficiency of electrolysis.

\end{abstract}



\section{Introduction}
\label{sec:intro}

Electrochemical processes, such as water electrolysis for hydrogen production, are a focus of efforts to develop a clean and efficient energy system.
Despite progress in advancing the catalytic properties of electrode materials, the efficiency of water splitting remains affected by the gas bubbles forming and growing
at the electrodes, which reduce the reaction area and impede mass transfer,
causing additional energy losses \citep{li2023bubble}. 

There is increased interest in structuring electrodes with regular surface elevations or alternating materials, particularly at the nano- and micrometer scales,
as this may accelerate bubble detachment 
  via reduced adhesion forces or fostered coalescence of neighboring bubbles \citep{li2023bubble,
bashkatov2024performance}. Additionally, small catalytic islands also reduce the expenditure of noble metal catalysts.
However, due to resolution limits of optical methods, it is difficult to study the behavior of small bubbles in experiments.
Nano-bubbles are typically observed indirectly through electrical signals \citep{chen2015JACS} or light scattering \citep{suvira2023imaging}, and distinguishing them from other surface adsorbates by atomic force microscopy remains challenging.

Molecular dynamics (MD) simulations have been successfully applied to advance the understanding of electrochemical nano-bubbles \citep{gadea2020electrochemically, ma2021dynamic}.
Also the stability theory for surface nano-bubbles developed in the last decade \citep{lohse2015surface} delivered valuable insights into the bubble evolution at nano-/micro-electrodes. 
In particular, it was clarified that contact line pinning in an oversaturated liquid leads to the stabilization of surface nano-bubbles in which the Laplace pressure is large \mbox{\citep{liu2014unified, PhysRevE.91.031003}}.
On the contrary, unpinned nano-bubbles, depending on oversaturation, will either dissolve or grow unlimitedly
but not stay stable \citep{PhysRevE.91.031003}.
This theory has been extended to electrochemical bubbles pinned at nano-electrodes
where influx due to electrochemically-generated gas may compensate the outflux due to Laplace pressure \citep{liu2017dynamic}.
\citet{zhang_lohse_2023} considered a \textit{reaction-controlled} growth mode by assuming that the bubble almost completely covers the electrode, so that all the gas produced at the electrode directly enters the bubble, and the electrolyte remains at zero-oversaturation. 
Recently, also a \textit{diffusion-controlled} mode was 
considered \citep{zhang2024threshold}, where the wetted electrode area is relatively large compared to the bubble size. Then, diffusion into the bulk becomes important. Assuming a linear concentration profile, the over-saturation can be derived from the current density.
For both modes mentioned, 
a minimum current density to stabilize the nano-bubbles was deduced.
\begin{figure}
 \centerline{
 \includegraphics[height=3.3cm]{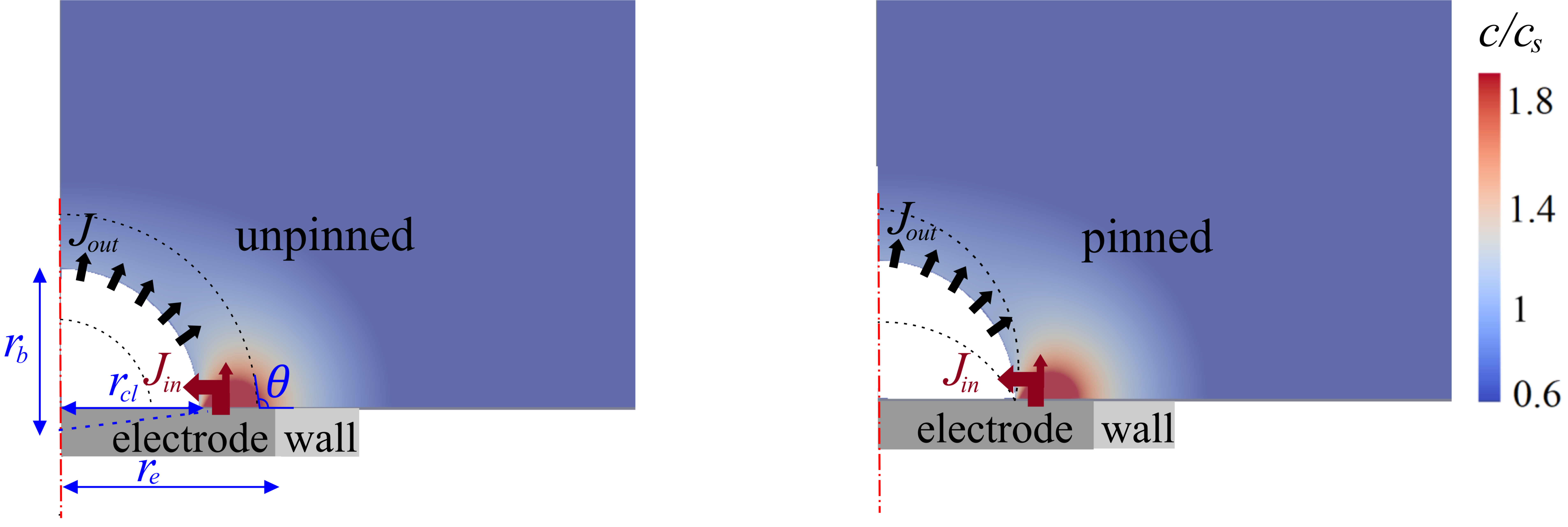} 
  }
  \captionsetup{justification=justified, width=\textwidth}
  \caption{Sketch of the unpinned and pinned bubble evolution process.
  The color surface represents the concentration ratio of dissolved gas $c$
  and saturation concentration $c_s$.
  }
\label{fig:sketch}
\end{figure}

In addition to nano-electrodes, bubbles evolving at micro-electrodes are interesting for practical reasons and have been studied intensively in e.g. \citep{bashkatov2022growth, park2023solutal}. However, the question of bubble stability still awaits a detailed investigation, which numerically is beyond the scope of MD.
Unlike nano-electrodes where growing bubbles early get pinned, micro-electrodes may lead to a more complex wetting behavior, including both pinned and unpinned manners \citep{yang2015dynamics, demirkir2024life}, see figure~\ref{fig:sketch}.
Depending on the bubble size compared to the wetted electrode area, the gas produced at the electrode may not completely enter the bubble. Parts of the produced gas could diffuse into the surrounding liquid, not directly contributing to bubble growth and creating a local over-saturated region near the bubble foot.
This is different from the idealized reaction- or diffusion-controlled modes
considered before at the nano-scale \citep{zhang_lohse_2023,zhang2024threshold} and requires an extension of the theory that will be presented below.

Therefore, this work aims at studying the dynamics and stability of 
both, pinned and unpinned,
hydrogen (H$_2$) bubbles  at micro-electrodes 
during water electrolysis, thereby accurately addressing the complex situation of the spatially inhomogeneous distribution of dissolved gas. This will be achieved by performing direct
numerical simulations, based on which the theory will be extended beyond pinned nano-bubbles.
Finally, this enables us to identify the parameter regions of bubble growth, dissolution and dynamic equilibrium and to demonstrate the stability of equilibrium states. 

\section{Numerical modelling}\label{sec:NumMet}
The gas-liquid interface is resolved using a geometric Volume-of-Fluid (VOF) method in {\it{Basilisk}}
\citep{popinet2023basilisk}.
The Navier–Stokes equations and the transport equation for the dissolved gas are solved with source terms to account for the phase change \citep{gennari2022phase}. 
To compute the surface tension, a height function method combined with a balanced-force discretization scheme \citep{popinet2009accurate} is used. The contact angle at the electrode surface is specified by the height function in the surface mesh cells \citep{afkhami2008height}.

An axisymmetric computational domain with a side length of 10 $r_{b,ini}$ is 
used, with initial bubble radius $r_{b,ini} = 5 - 50$ $\upmu$m. The electrode radius $r_e$ ranges from 5.5 to 100 $\upmu$m. 
Different constant current densities of $j=2.5-1250$ A/m$^2$ are applied to the wetted part of the electrode surface 
to resemble a potentiostatic operation mode, where the cell current reduces if the electrode blockage by bubbles increases.
According to Faraday's law, it yields corresponding Neumann boundary conditions of the concentration of dissolved H$_2$ ($c$) at the wetted electrode part.
At the remaining bottom wall, see figure~\ref{fig:sketch}, a no-flux condition is applied to the dissolved gas concentration.
For unpinned bubbles, static contact angles $\theta$ ranging from 45\textdegree{} to 120\textdegree{} are imposed. As we focus on the initial growth and stability of bubbles, the rapid change of $\theta$ shortly before detachment is not considered.
For the pinning cases considered, the bubble coverage on the electrode varies between 45\% $-$ 90\%. 
To keep the bubble pinned, we apply sufficiently large (150\textdegree) and small (30\textdegree) contact angles inside and outside the pinning point, respectively \citep{sakakeeny2021model}. 
Considering the slow or no motion of the contact line, 
a no-slip condition is used at the bottom boundary, which is validated in figure~\ref{fig:FS} in the Appendix.
Initially, we set the flow velocity to zero and the concentration to the bulk value $c_{b}$.
As often the initial hydrogen concentration in the bulk can be neglected compared to that at the bubble interface \citep{van2017electrolysis, gadea2020electrochemically}, we consider under-saturated/saturated electrolytes.
With $c_s$ denoting the saturation concentration at given external pressure, the over-saturation follows $\zeta=c_{b}/c_{s}-1 \leq 0$. If not stated otherwise, 
we consider
$c_b=0$, and thus $\zeta =-1$.
The material parameters used in the simulations, see table~\ref{tab:kd}, apply to water electrolysis in an aqueous electrolyte at standard conditions, except that the
diffusion coefficient is manually increased to accelerate the simulations. By rescaling the time according to the ratio of the increased to the real diffusion coefficient, the original bubble evolution is known to be recovered for the conditions considered in this work \citep{han2024}.
\begin{table}
  \begin{center}
\def~{\hphantom{0}}
  \begin{tabular}{lcc}
     Variables  & Value   &   Description   \\[3pt]
       $P_0$   & 100000 Pa & External pressure \\
       $c_{s}$   & 0.757 $mol/m^3$ & saturation concentration at $P_0$\\
       $z$  & 2 & charge number of reaction\\
       $D$   & $5 \cdot 10^{-9} \rightarrow 2 \cdot 10^{-5}$ & diffusion coefficient\\
       $M_g$ & 2 g/mol & molar mass of the hydrogen gas \\
       $\rho_g, \rho_l$ & 0.08, 1000 kg/m$^3$ & density of gas and liquid \\
       $\mu_g, \rho_l$ & $8.8 \cdot 10^{-6}$, $10^{-3}$ pa$\cdot$s & viscosity of gas and liquid \\
  \end{tabular}
  \caption{Material properties used in the simulations.}
  \label{tab:kd}
  \end{center}
\end{table}

\section{Results and Discussions}\label{sec:Res}
\begin{figure}
 \centerline{\includegraphics[height=5.65cm]{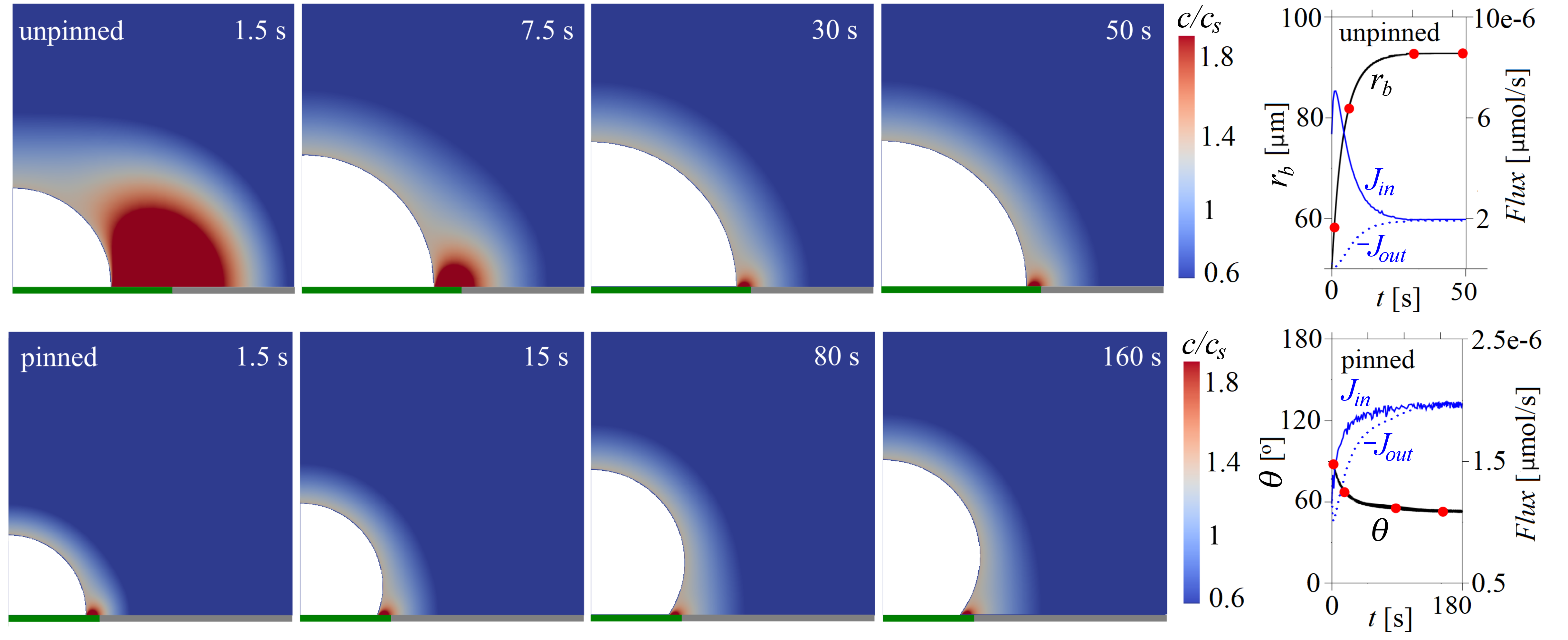}
 }
  \captionsetup{justification=justified, width=\textwidth}
  \caption{Left: Numerical results of the evolution of an unpinned (top) and a pinned (bottom) bubble. The color surface represents the distribution of dissolved gas concentration normalized by the saturation concentration $c_s$, the green bottom line marks the electrode.
  Right: Evolution of $r_b $ and $\theta$  with time.  The time instants shown 
  left are marked by red dots in the right graphs. For the unpinned case: $r_e=100\, \upmu$m, $j=125$ A/m$^2$, $\theta = 90$\textdegree, $r_{b,ini}=50\, \upmu$m.
  For the pinned case: $r_e=55\, \upmu$m,  $j=250$ A/m$^2$, $r_{cl}=50 \, \upmu$m, $\theta_{ini}=90$\textdegree{}. 
  }
\label{fig:BubEvo}
\end{figure}

Figure~\ref{fig:BubEvo} shows simulation results of how an unpinned and a pinned bubble develop over time.
Due to the electrode reaction, a high concentration region of dissolved H$_2$ near the wetted electrode part (red-color region) is clearly visible. The growth of the unpinned bubble (top) increases the electrode blockage. This reduces the amount of gas diffusing into the liquid, and the
high-concentration
region decreases in size. 
As shown in figure~\mbox{\ref{fig:sketch}}, this reduces the gas influx $ J_{in}$ [mol/s]
through the bubble interface near the bottom, where the liquid is over-saturated. In contrast, the gas outflux (magnitude $\left| J_{out} \right|$ [mol/s]) into the under-saturated bulk liquid increases with the expanding bubble surface area. 
Note that values of $J_{out}$ are negative, as defined later in \mbox{equation~\ref{eqn:Jout}}.
As shown in the right sub-figure, both fluxes counterbalance at about 30 s, at which the bubble reaches a dynamic equilibrium state, visible by the levelling-off of its radius. For the case of a pinned contact line (bottom), the contact angle reduces with time until an equilibrium is reached after 80 s. Although the wetted electrode area remains constant, the high-$c_\mathrm{H_2}$ region seems to slightly diminish, as the smaller contact angle also reduces gas transport into the bulk. The right sub-figure shows that this leads to an increase in $ J_{in} $ with time, while $\left| J_{out} \right| $ also rises with the growing bubble surface area, until both converge. 
Therefore, a dynamic equilibrium is found for both unpinned and pinned bubbles at micro-electrodes, which will be further analyzed below.

\subsection{Theoretical analysis}\label{sec:TheoAnal}
\begin{figure}
 \centerline{
 \includegraphics[height=4.5cm]{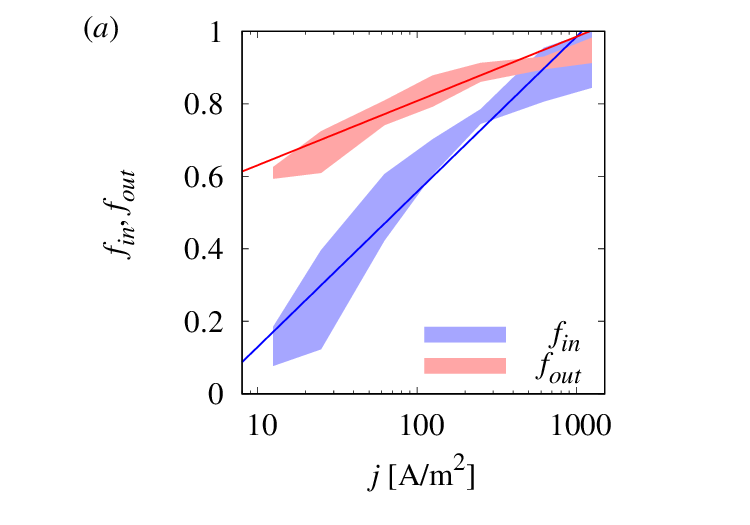} \ \ \ \ \ 
  \includegraphics[height=4.5cm]{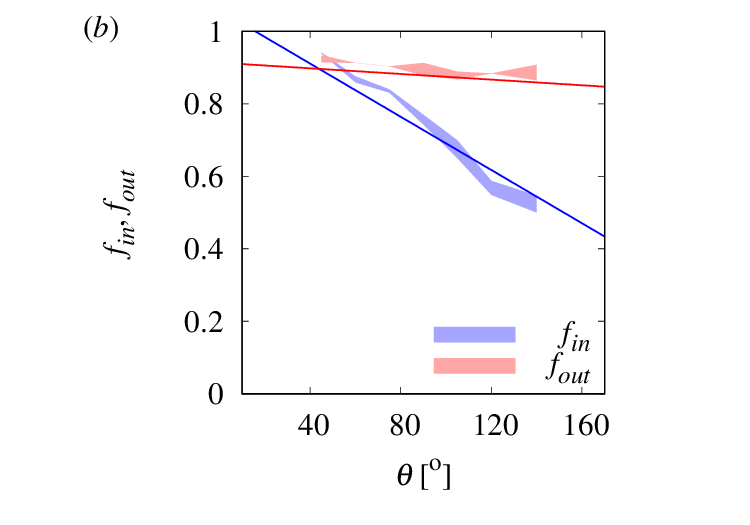}
  }
  \captionsetup{justification=justified, width=\textwidth}
  \caption{
$f_{in}$ and $f_{out}$ derived from numerical data when unpinned bubbles reach an equilibrium in a liquid of $\zeta = -1$, i.e. $c_b=0$. 
(\textit{a}) Influence of the applied current density. $r_{b,ini}=50 \, \upmu$m, $\theta=90$\textdegree{}, 
color region represents data range obtained by $r_e=50-150\,  \upmu$m.
(\textit{b}) Influence of the contact angle. $j=250$ A/m$^2$,
color region represents data range obtained by  $r_{b,ini}=50 \, \upmu$m, $r_e = 150\,  \upmu $m or $r_{b,ini}=100\,  \upmu$m, $r_e = 125\,  \upmu $m. 
Solid lines represent fitting functions (equations~\ref{eqn:finjtheta}, \ref{eqn:foutjtheta} in Appendix) used in the theoretical solutions.
}
\label{fig:finfout}
\end{figure}

As evidenced by the red high-concentration region near the bubble foot in figure~\ref{fig:BubEvo}, even when the bubble already covers most parts of the electrode, 
gas diffusion into the surrounding electrolyte takes place.
Therefore, we start from the reaction-controlled modelling \citep{zhang_lohse_2023}, which calculates the gas entering the bubble directly from the current density $j$,
but introduce a correction factor $f_{in}<1$ to account for gas remaining in the electrolyte.
The gas influx $J_{in}$ across the bubble surface can then be described as follows: 

\begin{equation}
J_{in} = f_{in} \cdot J_{e} = f_{in} \cdot \dfrac{  j \pi (r_e^2 - r_{cl}^2)}{ z F},
    \label{eqn:Jin}
\end{equation}
with $J_e, \, r_e, \, r_{cl}, \, z, \, F$ denoting the gas flux generated at the electrode, the radius of the electrode and the bubble contact line, the charge number, and the Faraday constant. 
Values of $f_{in}$ are derived from numerical simulations 
for different reaction conditions
by computing the ratio of the gas-flux that actually enters the bubble and that produced at the electrode.
As shown in figure~\ref{fig:finfout}, $f_{in}$ increases towards 1 with 
enhancing $j$, where the bubbles grow larger and the gas loss into the bulk reduces. At more hydrophilic surfaces, the bubble shape will increasingly impede gas transport towards the bulk, figure~\ref{fig:BubEvo}, thus causing also a higher $f_{in}$. 

When the surrounding liquid is under-/saturated ($\zeta \leq 0$),
at the same time, gas may diffuse out of the bubble. Assuming that convection effects are negligibly 
small, the gas transport equation can be simplified for steady states to $\nabla^2 c = 0$.
Combining this with Fick's and Henry's law \citep{popov2005evaporative, PhysRevE.91.031003}, 
the gas outflux $J_{out}$ reads:
\begin{equation}
J_{out} = 
f_{out} \cdot \pi r_b D  c_{s}\left( \zeta - \dfrac{2 \gamma}{r_b P_0} \right) 
f_p \sin \theta .
    \label{eqn:Jout}
\end{equation}
Here, $r_b, \, D, \,  \gamma $ denote the bubble radius, the gas diffusion coefficient, and the surface tension. The shape factor $f_p$ introduced by \cite{popov2005evaporative}, see Appendix~\ref{sec:finoutp}, depends only on the water-side contact angle $\theta$, and monotonically decreases from infinity towards unity when $\theta$ increases from 0\textdegree{} to 180\textdegree{}.
In addition to previous work, the factor $f_{out}\le 1$ is introduced in equation~\ref{eqn:Jout}. 
It accounts for reduced outflux due to a high-concentration region appearing near the bubble foot from the electrode reaction (figure~\ref{fig:BubEvo}).
In more detail, 
it is defined as
the fraction of the bubble surface area where gas leaves the bubble.
Without oversaturated regions
around the bubble, $f_{out} = 1$.
The behavior of $f_{out}$ determined from numerical simulations is shown in figure~\ref{fig:finfout},
where the colored regions mark the results obtained 
for various configurations given in the caption.
Similarly to $f_{in}$, $f_{out}$ increases with  $j$ as the bubbles become larger. The contact angle seems to have only a minor influence. Note that 
for the different configurations,
only small 
variations in
$f_{in}$ and $f_{out}$ are observed, which motivates us to approximate 
both
by fitting functions (solid lines), as detailed in Appendix~\ref{sec:finoutp}.

Mass transfer at the gas-liquid interface determines how the bubble geometry evolves with time.
Applying the ideal gas law, the time derivative of the bubble volume can be obtained:
$dV/dt = M_g/\rho_g \left( J_{in} + J_{out} \right)$,  with $M_g, \, \rho_g$ denoting the molar mass and the density of the gas.
Depending on the reaction parameters, the bubble could grow ($ J_{in}+  J_{out}>0$) or shrink ($J_{in}+  J_{out} < 0$) with time.
If the shape of the cap remains spherical,
an unpinned bubble evolving with a constant 
contact angle can be described as follows:
\begin{equation}
     \frac{d r_\mathrm{b}}{dt} = 
   \dfrac{M_g \left[ f_{in} \cdot \dfrac{  j  (r_e^2 - r_{cl}^2)}{ z F} + f_{out} \cdot r_b D  c_{s}\left( \zeta - \dfrac{2 \gamma}{r_b P_0} \right) f_p
   \sin \theta \right]}{\rho_g r_b^2 ( 2 + 3 \cos\theta - \cos^3\theta  ) } .
\label{eqn:drt2}
\end{equation}
For the case of a pinned bubble, the temporal change of the contact angle is derived as:
\begin{equation}
    \frac{d \theta}{dt} = 
       \dfrac{M_g \left[ -f_{in} \cdot \dfrac{  j  (r_e^2 - r_{cl}^2)}{ z F} - f_{out} \cdot r_{cl} D  c_{s}\left( \zeta - \dfrac{2 \sin\theta \gamma}{r_{cl} P_0} \right)  f_p \right](1-\mathrm{cos}\theta )^2}{\rho_g r_{cl}^3  } .
       \label{eqn:dthetat2}
\end{equation}
If during the evolution of the bubble, a counterbalance between $J_{in}$ and $J_{out}$ is reached, as shown in figure~\ref{fig:BubEvo}, a dynamic equilibrium
is found. From equations~\ref{eqn:Jin}, \ref{eqn:Jout}, it can be derived:
\begin{equation}
   \left[ f_{in} \dfrac{  j  (r_e^2 - r_{cl}^2)}{ z F} + f_{out} r_{b} D  c_{s}\left( \zeta - \dfrac{2 \gamma}{r_{b} P_0} \right) f_p \sin \theta \right]_{eq} = 0.
    \label{eqn:eq}
\end{equation}
Subscript \textit{eq} denotes \textit{equilibrium}.
Equation~\ref{eqn:eq} allows to determine $r_{b,eq}$ or $\theta_{eq}$ for unpinned or pinned bubbles, respectively. If no root can be found for the given conditions, an equilibrium will not occur, and the bubble either grows unlimitedly or completely dissolves.

\begin{table}
  \begin{center}
\def~{\hphantom{0}}
  \begin{tabular}{lcc}
    Influence factors  &Unpinned bubbles   &   Pinned bubbles   \\[3pt]
    Bubble surface area   & stabilizing & stabilizing \\
    Wetted electrode area   &  stabilizing &  no change\\
     Laplace pressure  & destabilizing & $\theta >90$\textdegree{}: stabilizing; \ \ $\theta <90$\textdegree{}: destabilizing \\
     Contact angle  & no change & destabilizing  \\
  \end{tabular}
  \caption{
  Stabilizing and destabilizing factors for bubbles in an under-/saturated liquid 
  }
  \label{tab:stab}
  \end{center}
\end{table}
Once a dynamic equilibrium is reached, minor disturbances in the system may arise.
Whether these disturbances will grow or diminish determines the \textit{stability} of the equilibrium.
\cite{PhysRevE.91.031003} studied surface nano-bubbles 
in oversaturated water, indicating that the equilibrium is
 \textit{stable} for pinned and
\textit{unstable} for unpinned bubbles.
In electrolysis, the gas production at the electrode surface may lead to a different stability behavior.
For example, if an unpinned bubble becomes temporarily larger than $r_{b,eq}$, 
$J_{in}$ will decrease due to the reduced wetted electrode area, whereas the bubble surface area and therefore $\left| J_{out} \right|$ increase. 
Thus, both changes tend to bring the bubble back to $r_{b,eq}$.
Other factors influencing the stability are analyzed 
in table \ref{tab:stab}.
As it is difficult to decide which is the determining factor, 
we investigate the stability
criteria for unpinned and pinned
bubbles
\citep{PhysRevE.91.031003}:

\begin{equation}\label{eqn:stab0}
  \left[\dfrac{ \partial }{\partial r_b}\frac{dr_b}{dt}\right]_{eq} < 0, \, \, \, \, \, \left[\dfrac{ \partial }{\partial \theta} \frac{d\theta}{dt}\right]_{eq} < 0.
\end{equation}
For unpinned surface bubbles, combining equations \ref{eqn:drt2} and \ref{eqn:eq}, we obtain (derivation in Appendix~\ref{sec:deri}):
\begin{equation}\label{eqn:drtr}
   \left[ \dfrac{ \partial }{\partial r_b} \frac{dr_b}{dt} \right]_{eq} =\dfrac{M_g }{\rho_g  ( 2 + 3 \cos\theta - \cos^3\theta  ) r_{b,eq}^2} \left[
    \zeta P_0 K_2 - \dfrac{4K_1 r_{b,eq} \sin^2 \theta}{r_e^2}
    \right],
\end{equation}
with $K_1, \, K_2$ being positive constants at given values of $j, \, r_e$ and $\theta$:
\begin{equation}
    K_1 = \dfrac{f_{in}jr_e^2}{2zF}, \, \, \, \, \, K_2 = \dfrac{f_{out}D c_{s} f_p \sin \theta }{P_0}.
    \label{Aeqn:K1K20}
\end{equation}
Because $ (2 + 3 \cos\theta - \cos^3\theta ) >0,$ 
the sign of expression
\ref{eqn:drtr} is determined by the sign of the part in the
square brackets on the right hand side.
For the case of $\zeta \leq 0$, this sign is always negative, indicating a stable equilibrium for unpinned bubbles in under-/saturated liquids.
However, for larger $\zeta$, the sign is likely to become positive, in agreement with the statement in \citet{PhysRevE.91.031003} that unpinned bubbles are unstable in over-saturated liquids.

For pinned surface bubbles, combining equations \ref{eqn:dthetat2} and \ref{eqn:eq} leads to (derivation in Appendix~\ref{sec:deri}):
\begin{equation}\label{eqn:dthetattheta2}
    \left[\dfrac{ \partial }{\partial \theta} \dfrac{d\theta}{dt}\right]_{eq} = \dfrac{M_g }{\rho_g  r_{cl}^2 } (1-\cos \theta_{eq})^2 \left[
     \dfrac{2K_4 \gamma}{r_{cl}}f_p \cos\theta + 
     K_3 \left(\dfrac{f_{in}}{f_p}\frac{df_p}{d\theta} - \frac{\partial f_{in}}{ \partial \theta}
     \right)
    \right]_{eq},
\end{equation}
with $K_3, \, K_4$ being positive constants at given values
of $j, \, r_e$ and $r_{cl}$:
\begin{equation}
    K_3 = \dfrac{ j  (r_e^2 - r_{cl}^2)}{zFr_{cl}} , \, \, \, \, \, 
    K_4 = \dfrac{f_{out}D c_{s}  }{P_0}.
    \label{Aeqn:K3K40}
\end{equation}
The sign of expression \ref{eqn:dthetattheta2} is determined by the sign of the part in the square brackets on the right hand side.
As will be discussed below 
(figure~\ref{fig:dynEqStPin}),
for small bubbles ($r_{cl} \sim 1 \, \upmu$m) the sign is negative when $\theta > 90$\textdegree{} and positive when $\theta < 90$\textdegree{},
indicating that the sign is mainly determined by the first term inside the bracket.
Thus, only pinned small bubbles of flat shape tend to be stable, while taller caps get unstable by the change of the Laplace pressure (table~\ref{tab:stab}).
This generalizes the finding in \citet{PhysRevE.91.031003} that pinned nanobubbles, which are typically flat, are stable.
For larger bubbles, the second term in the bracket may become dominant.
Although both, $df_p/d\theta$ and $\partial f_{in} / \partial \theta$ are negative, our calculations reveal that the sign is generally negative for bubbles of $\sim 50$ $\upmu$m in size.
This corresponds to the fact that $f_p$ changes 
faster than $f_{in}$ with $\theta$, as can be seen in Appendix~\ref{sec:finoutp}.		
Therefore, we conclude here that 
in under-saturated/saturated bulk electrolytes, bubbles evolving on micro-electrodes in either pinned or unpinned mode may reach a stable equilibrium state.
But we remark that under certain conditions, e.g. at large current \citep{zhang_lohse_2023}, the equilibrium state ($J_{in}+J_{out}=0$) may not be achieved, letting alone its stability.

\subsection{Numerical simulations}
\begin{figure}
 \centerline{
 \includegraphics[height=4.5cm]{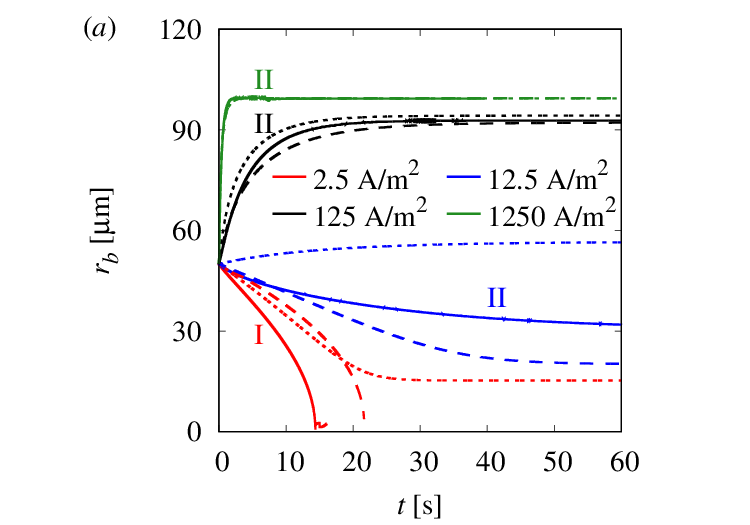}\ \ \ \ \ \
 \includegraphics[height=4.5cm]{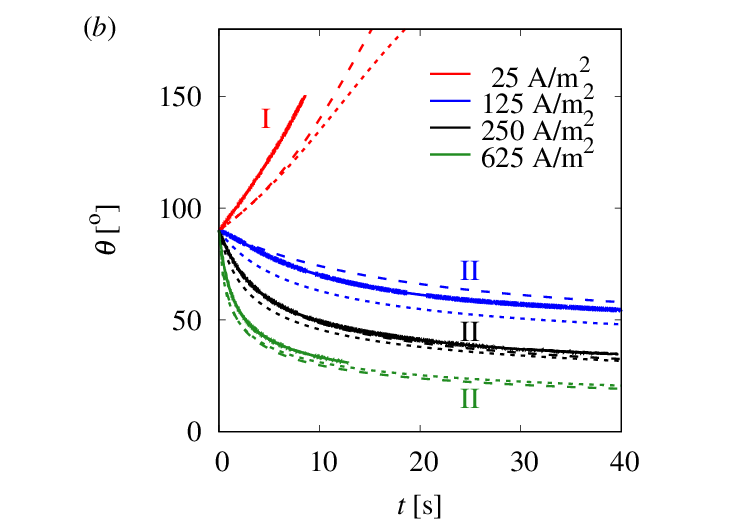}
  }
\captionsetup{justification=justified, width=\textwidth}
  \caption{(\textit{a}) Evolution of the radius of an unpinned bubble, $\theta = 90$\textdegree, $r_e=100\, \upmu$m, $r_{b,ini}=50 \, \upmu$m.
  (\textit{b}) Evolution of the contact angle of a pinned bubble, $r_{cl}=44 \, \upmu$m, $r_e=55\, \upmu$m, $\theta_{ini}=90$\textdegree. 
  Regime I: bubble dissolves completely.
  Regime II: bubble reaches a dynamic equilibrium.
  Solid lines: simulation. 
  Dashed lines: theoretical solution with $f_{in/out}$.
  Dotted lines: theoretical solution without $f_{in/out}$. 
  Note that the simulations are stopped when $\theta >$ 150\textdegree{} or $\theta <$ 30\textdegree{} to avoid numerical difficulties in interface reconstruction.
  }
\label{fig:bubDyn}
\end{figure}

In the following, we investigate the conditions of equilibrium and stability in more detail by combining numerical simulations and theoretical reasoning.
Figure~\ref{fig:bubDyn} shows numerical and theoretical results of bubble evolution with (dashed lines) and without (dotted lines) the consideration of gas loss into the electrolyte ($f_{in}, \, f_{out}$),
equations~\ref{eqn:drt2} and \ref{eqn:dthetat2}.
When the applied current density increases, the bubble evolution tends to change from complete dissolution (Regime I) towards dynamic equilibrium states (Regime II).
Unlimited growth until detachment is expected to occur at larger $r_b$ and smaller $\theta$, which is out of the scope of our simulations.
We remark that the bubble end state is not influenced by the initial state,
see Appendix~\ref{sec:Indephy}.
This indicates that convection is not important ($Pe=r_b/D \cdot dr_b/dt \ll 1$), and that bubbles temporarily at different states all tend to converge to the equilibrium state, i.e.~the stability condition (\ref{eqn:stab0}) is fulfilled.
Because $f_{in}$ is in general smaller than $f_{out}$ (figure~\ref{fig:finfout}), 
adding them to the original theoretical solution of \citet{zhang_lohse_2023} causes slower bubble growth and faster dissolution. This improves the agreement with the numerical simulations especially at smaller currents.

\begin{figure}
 \centerline{
  \includegraphics[height=4.5cm]{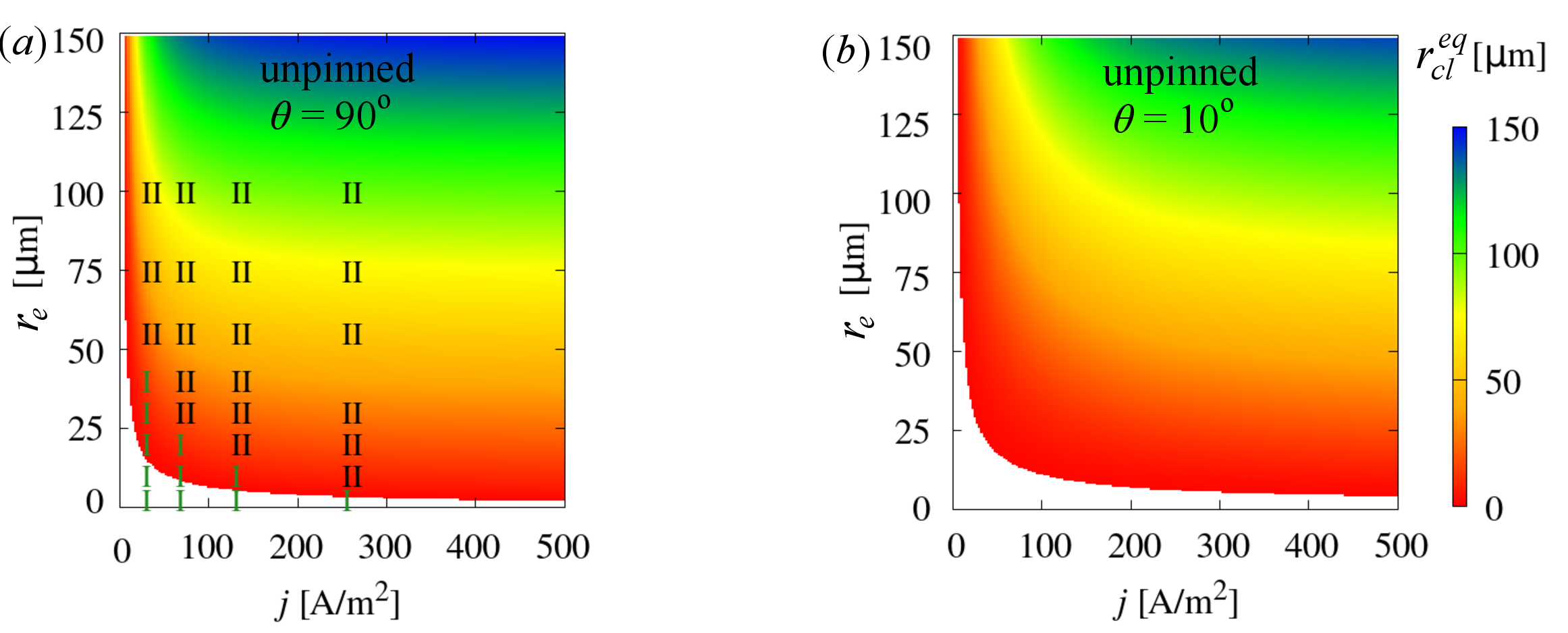}
  }
 \captionsetup{justification=justified, width=\textwidth}
  \caption{Influence of electrode radius $r_e$ and current density $j$ on the equilibrium contact line radius $r_{cl}^{eq}$ of unpinned bubbles when the contact angle is (\textit{a}) 90\textdegree{} and (\textit{b}) 10\textdegree{}. 
  Color surface: Theoretical solution. 
  White area represents complete dissolution.
  Numerical results of bubble end states
  (I: complete dissolution; II: dynamic equilibrium)
  are added in (\textit{a}).
  }
\label{fig:dynEqStExp}
\end{figure}

Next, by using the adapted theoretical solution, 
equation~\ref{eqn:eq},
we provide a systematic view of how the final bubble state is influenced by current density, electrode size and wettability.
The color surface in figure~\ref{fig:dynEqStExp}(\textit{a})
represents the contact line radius for unpinned bubbles with a contact angle of 90\textdegree{} when a dynamic equilibrium is reached (Regime II).
In the white space below the color region, no positive root of equation~(\ref{eqn:eq}) exists, i.e.~the bubble completely dissolves (Regime I).
The bubble 
end states obtained by numerical simulations are added 
and are found
to qualitatively reproduce the
theoretical results.
As can be seen, the equilibrium bubble size decreases 
when lowering $r_e$ and $j$.
For each current density, there exists a critical electrode size below which bubbles become unstable on the electrode,
which decreases at larger $j$.
The critical $r_e$ obtained from the simulations is only slightly larger than 
for the theoretical solutions.
For more hydrophilic electrodes, as shown in figure~\ref{fig:dynEqStExp}(\textit{b}) for the case of $\theta=10$\textdegree{}, $r_{cl,eq}$ 
becomes smaller, and the dissolution region extends.
Note that the results shown here are for the case $\zeta=-1$. 
However, a similar behavior of the bubble evolution was found also for $\zeta=0$, see Appendix~\ref{sec:Indebulk}.

\begin{figure}
 \centerline{
  \includegraphics[height=4.5cm]{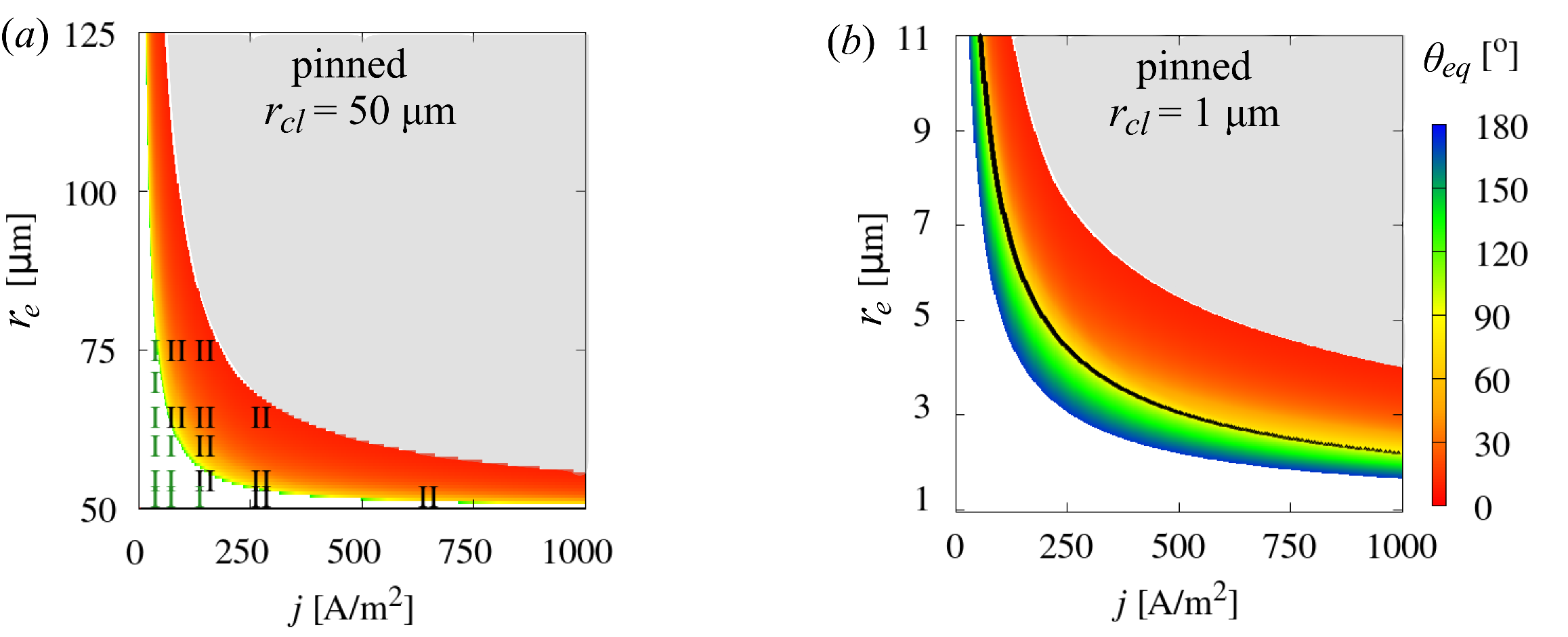}
  }
 \captionsetup{justification=justified, width=\textwidth}
  \caption{
Influence of electrode radius $r_e$ and current density $j$ on the equilibrium contact angle $\theta_{eq}$ of pinned bubbles with (\textit{a})~$r_{cl}=50\,\upmu$m or (\textit{b})~$r_{cl}=1\,\upmu$m. 
  Color surface: Theoretical solution. 
  White or grey areas represent complete dissolution or unlimited growth.
  Numerical results of bubble end states
  (I: complete dissolution; II: dynamic equilibrium)
  are added in (\textit{a}).
  The black curve in (\textit{b}) marks $ \left[ \partial (d\theta / dt) / \partial \theta \right]_{eq} =0$,
  which is not visible in (\textit{a}) where the
  expression is
  always negative.
  }
\label{fig:dynEqStPin}
\end{figure}

Figure~\ref{fig:dynEqStPin} shows the final contact angle of pinned bubbles with a pinning radius of 50~$\upmu$m~(\textit{a}) and 1~$\upmu$m~(\textit{b}).
The white space below the color region represents the cases of complete dissolution ($\theta \rightarrow 180$\textdegree{}), and the upper grey area represents unlimited growth ($\theta \rightarrow 0$\textdegree{}).
As can further be seen, smaller $r_e$ and lower $j$ lead to more flat bubbles.
Such a change in bubble shape reduces the surface available for $J_{out}$, thus to maintain the equilibrium at higher Laplace pressure and slower gas production.
For each current density, also here exists a critical $r_e$ below which bubbles always dissolve, which decreases with increasing $j$.

A bubble does not necessarily stay stable after an equilibrium state 
has temporarily been reached.
As mentioned above, 
unpinned bubbles are always stable in liquids of $\zeta \leq 0$ (equation~\ref{eqn:drtr}).
For pinned bubbles of $r_{cl}=50\, \upmu$m, the sign of $ \left[\frac{ \partial }{\partial \theta} \frac{d\theta}{dt}\right]_{eq} $ (equation~\ref{eqn:dthetattheta2}) is found to be negative, i.e. the bubbles have a stable equilibrium.
For smaller bubbles of $r_{cl}=1\, \upmu$m,
we plot the isoline $ \left[\frac{ \partial }{\partial \theta} \frac{d\theta}{dt}\right]_{eq} =0$ 
as black solid line in figure~\ref{fig:dynEqStExp}(\textit{b}).
We find that the sign 
of this expression is negative 
for $\theta>90$\textdegree{} and positive for $\theta < 90$\textdegree{} in general.
This confirms the analysis below equation~\ref{Aeqn:K3K40}.
More details on the distribution of the sign can be found in 
Appendix~\ref{sec:deri}.

\subsection{Impact on reaction efficiency}\label{sec:Effi}
\begin{figure}
 \centerline{
   \includegraphics[height=5cm]{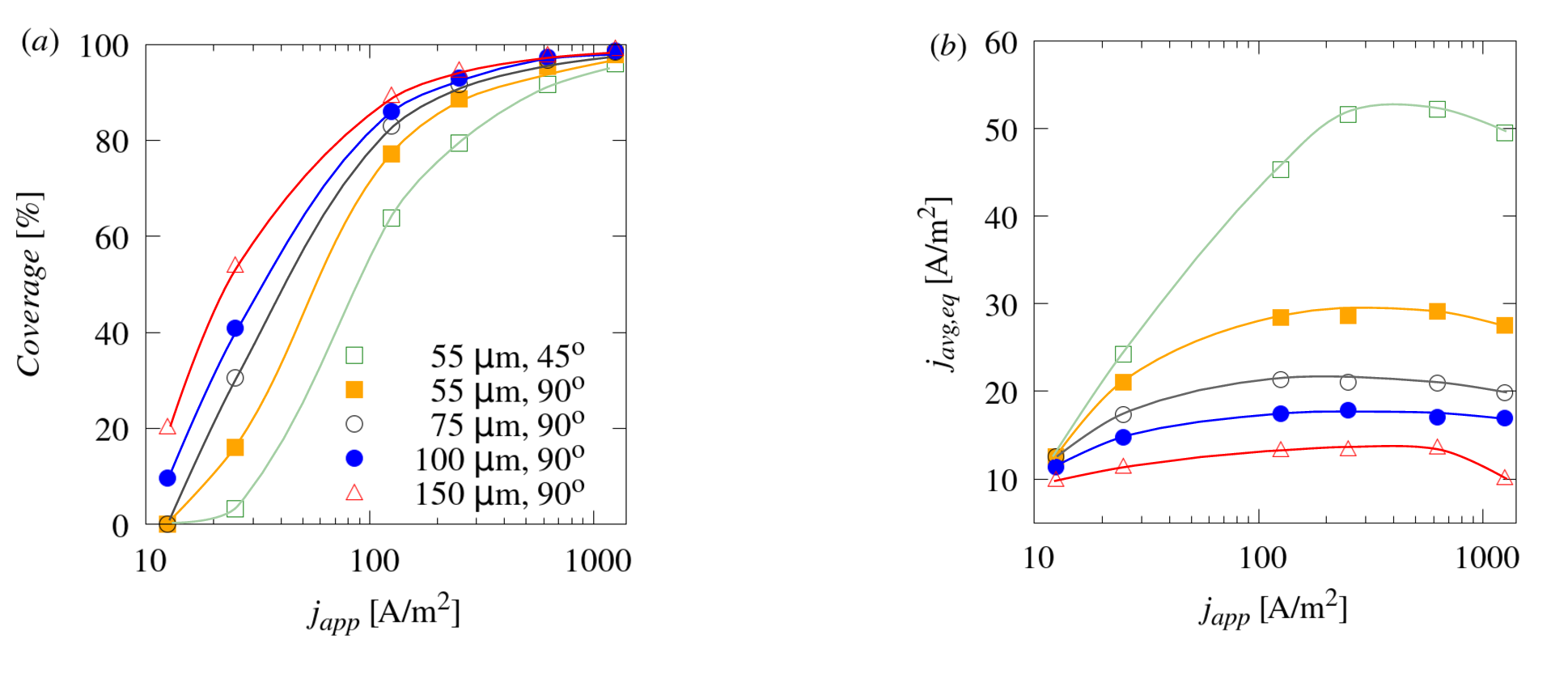}
  }
\captionsetup{justification=justified, width=\textwidth}
  \caption{(\textit{a}) Electrode coverage and (\textit{b}) current density averaged over the whole electrode surface area for unpinned bubbles at 
 equilibrium for micro-electrodes of different size and wettability, $r_{b,ini}=50 \, \upmu $m.}
\label{fig:Coveragejavg}
\end{figure}

Finally, we discuss how to design the electrode to  
effectively regulate the bubble equilibrium, 
aiming to reduce electrode blockage 
and thus 
related energy losses.
For an unpinned bubble, its contact radius determines the active area of the electrode for reaction.
Figure~\ref{fig:Coveragejavg}(\textit{a}) shows how
the fraction of the electrode surface covered by the bubble changes with the current density applied ($j_{app}$). 
In general, lower $j_{app}$, smaller and more hydrophilic electrodes could reduce the bubble coverage.
Figure~\ref{fig:Coveragejavg}(\textit{b}) shows 
the current density averaged over the whole electrode area, $j_{avg}$, as a function of $j_{app}$, 
which sheds insight onto the experimentally measurable $current-potential$ curve \citep{bard2022electrochemical}: an indicator for the energy transfer efficiency of the electrolysis.
When $j_{app}$ increases, $j_{avg}$ first increases then levels off, even slightly decreases.
Smaller and hydrophilic electrodes are beneficial for enhancing the efficiency.
These trends correlate with lower bubble coverage on the electrode.
Further, when the electrode is increasingly blocked by the bubble, the electric current lines will concentrate near the electrode edge. This may lead to a temperature hotspot, causing a thermal Marangoni force that retards the bubble detachment \citep{hossain2020thermocapillary}.
If a bubble is pinned by geometrical or chemical surface heterogeneities, the bubble coverage of the electrode is pre-defined, and the reaction efficiency can only be influenced by the bubble shape.
For smaller $\theta$,
the electric current lines will be more distorted near the bubble, resulting in a stronger shielding effect and larger Ohmic loss.
As can be seen in figure~\mbox{\ref{fig:dynEqStExp}}, $\theta_{eq}$ increases when $r_e, r_{cl}$ and $j_{app}$ decrease.
This again indicates the advantage of using 
electrodes of smaller size
or electrodes structured with an array of nano-/micrometer sized catalytic spots.

\section{Conclusions}\label{sec:concF}
We studied the equilibrium and stability of gas bubbles evolving 
at micro-electrodes using direct numerical simulations and theoretical analysis by adapting an existing theoretical model for the reaction-controlled mode.
At micro-electrodes, both pinned and unpinned bubbles may reach a dynamic equilibrium state, at which the in- and outfluxes of gas across the bubble surface counterbalance each other.
In under- and saturated bulk electrolytes,
this equilibrium is always stable for unpinned bubbles.
Under contact line pinning, it is stable for micro-bubbles and flat nano-bubbles 
when the contact angle is larger than 90\textdegree{}.
Both numerical and theoretical solutions suggest a critical electrode size, below which formed bubbles at low oversaturation will eventually dissolve. 
This work 
gives support to recent
experimental observations of
stationary bubbles and bubble dissolution \mbox{\citep{suvira2023imaging}}. 
Our 
findings
may also contribute to future efficiency enhancements of water electrolysis
by 
using hydrophilic
nano-/microstructured
electrodes
at moderate current densities.


\backsection[Acknowledgements]{We gratefully acknowledge discussions with Changsheng Chen and Yunqing Ma. 
}

\backsection[Funding]{
Financial support by
the National Natural Science Foundation of China, 
grant no. 22309007,
the Federal Ministry of Education and Research within the project H2Giga-SINEWAVE, grant no. 03HY123E, 
the Graduate Academy of TU Dresden and the Professorinnenprogramm III of the Federal Government and the Länder
is greatly acknowledged. 
}

\backsection[Declaration of interests]{The authors report no conflict of interest.}


\backsection[Author ORCIDs]{Mengyuan Huang, https://orcid.org/0000-0002-4093-6992; Chao Sun, https://orcid.org/0000-0002-0930-6343; Kerstin Eckert, https://orcid.org/0000-0002-9671-8628; Xianren Zhang, https://orcid.org/0000-0002-8026-9012; Gerd Mutschke, https://orcid.org/0000-0002-7918-7474}


\appendix
\section{Details of expressions and  derivations}\label{appA}

\subsection{$f_{in}, \, f_{out}, \, f_p$}\label{sec:finoutp}
According to figure~\ref{fig:finfout}, only small variations can be observed when the electrode size changes.
Therefore, we
simplify the 
theoretical solutions by
using fitting functions
for 
 $f_{in}$ and $f_{out}$.
 These read:
\begin{equation}
    f_{in}(j,\theta) = 1.378 (-0.0037 \theta + 1.066)\cdot (0.184 ln(j) - 0.296).
\label{eqn:finjtheta}
\end{equation}
\begin{equation}
   f_{out}(j,\theta) = 1.139 (-0.0004\theta + 0.938)\cdot (0.075 ln(j) +0.441).
\label{eqn:foutjtheta}
\end{equation}

\begin{figure}
	\centering
	\includegraphics[height=5cm]{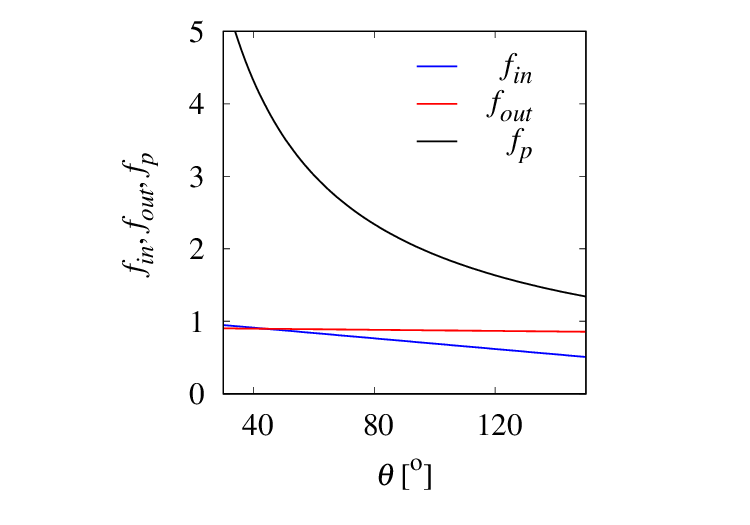}
	 \captionsetup{justification=justified, width=\textwidth}
	\caption{Comparison of $f_{in}, \, f_{out}$ ($j=250$ A/m$^2$) and $f_p$ depending on the water-side contact angle. 
	}
\label{fig:finoutp}		
\end{figure}

The shape factor
$f_p$
used to compute $J_{out}$ is defined as follows \citep{popov2005evaporative}:
\begin{equation}
f_p(\theta ) = \dfrac{\mathrm{sin} \theta}{1 - \mathrm{cos} \theta} + 4 \int_{0}^{\infty} \dfrac{1+\mathrm{cosh} [ 2 (\pi - \theta) \tau ] }{\mathrm{sinh} (2 \pi \tau)} \mathrm{tanh}(\theta \tau)\,d\tau,
\label{eqn:ftheta}
\end{equation}
with $\theta$ denoting the water-side contact angle.
It can be numerically
approximated
by the fitting function
$f_p(\theta)=112.55 \theta^{-0.884}$.
Figure~\ref{fig:finoutp} compares
the behavior of
$f_{in}, \, f_{out}$ and $f_p$ versus $\theta$. 
In general, all factors decrease with increased $\theta$, among which the Popov factor $f_p$ changes fastest.
The influence of $\theta$ on $f_{out}$ is only minor and is
neglected in the
derivation of bubble stability.

\subsection{Derivations of bubble stability}\label{sec:deri}
For the unpinned bubbles, based on equation~\ref{eqn:drt2}, the temporal change of the curvature radius can be expressed as:
\begin{align}
 \dfrac{dr_b}{dt} &=  
  \dfrac{M_g }{\rho_g  ( 2 + 3 \cos\theta - \cos^3\theta  ) } \left[
  \frac{1}{r_b^2}\left(
  \dfrac{f_{in}jr_e^2}{zF} - \dfrac{2 f_{out} \gamma D  c_{s}  \sin \theta f_p  }{P_0}
  \right) \right]\\
  &+ \dfrac{M_g }{\rho_g  ( 2 + 3 \cos\theta - \cos^3\theta  ) } \left[ \frac{1}{r_b}
  f_{out} D  c_{s}  \zeta \sin \theta f_p - \dfrac{f_{in}j \sin^2\theta}{zF}\right].
 \label{Aeqn:drt3}
\end{align}

Defining variables $K_1,\, K_2$ that are positive constant values during each bubble evolution event with constant $j, \, r_e$ and $\theta$,
\begin{equation}
    K_1 = \dfrac{f_{in}jr_e^2}{2zF}, \, \, \, \, \, K_2 = \dfrac{f_{out}D c_{s} \sin \theta f_p }{P_0},
    \label{Aeqn:K1K2}
\end{equation}

the partial derivative of equation~\ref{Aeqn:drt3} can be expressed as follows:
\begin{align}\label{Aeqn:drtr}
    \dfrac{ \partial }{\partial r_b} \frac{dr_b}{dt} &=\dfrac{M_g }{\rho_g  ( 2 + 3 \cos\theta - \cos^3\theta  ) } \dfrac{\partial}{\partial r_b}  \left[
    \frac{1}{r_b^2}(2K_1 - 2\gamma K_2) +  \frac{1}{r_b}(\zeta P_0 K_2)
    \right] \\
    &=\dfrac{M_g }{\rho_g  ( 2 + 3 \cos\theta - \cos^3\theta  ) r_b^2} \left[
    \frac{4(\gamma K_2 - K_1) - \zeta P_0 K_2 r_b}{r_b}
    \right]. \nonumber
\end{align}

At equilibrium, combining equations~\ref{eqn:eq} and \ref{Aeqn:K1K2}, we obtain:
\begin{equation}
    \gamma K_2 - K_1 = \frac{1}{2}\left( 
    \zeta P_0 K_2 r_{b,eq} - \dfrac{2K_1 r_{b,eq}^2 \sin^2 \theta}{r_e^2}
    \right).
\end{equation}

Then, equation~\ref{Aeqn:drtr} at equilibrium can be simplified as:
\begin{equation}
    \left[\dfrac{ \partial }{\partial r_b} \frac{dr_b}{dt} \right]_{eq}=\dfrac{M_g }{\rho_g  ( 2 + 3 \cos\theta - \cos^3\theta  ) r_{b,eq}^2} \left[
    \zeta P_0 K_2 - \dfrac{4K_1 r_{b,eq} \sin^2 \theta}{r_e^2}
    \right].
    \label{Aeqn:drtr2}
\end{equation}

Because $ (2 + 3 \cos\theta - \cos^3\theta ) >0$,
the sign of $\left[ \frac{\partial}{ \partial r_b} \frac{dr_b}{dt}\right]_{eq}$ is determined by the sign of the sqaure bracket on the right hand side of equation~\ref{Aeqn:drtr2}:
\begin{equation}
    \zeta P_0 K_2 - \dfrac{4K_1 r_{b,eq} \sin^2 \theta}{r_e^2},
    \label{eqn:drtrIndi}
\end{equation}
which is negative if $\zeta \leq 0$.
We remark that this expression can also be readily applied to bubbles on non-reactive surfaces by setting $K_1=0$, $f_{out}=1$.

For a pinned bubble, based on equation~\ref{eqn:dthetat2}, we express the temporal change of the water-side contact angle as follows:
\begin{align} \label{Aeqn:dthetat3}
 \dfrac{d\theta}{dt} &=  
  \dfrac{M_g }{\rho_g  r_{cl}^2 } \left[
 \left( \dfrac{ 2f_{out}Dc_{s}\gamma}{r_{cl}P_0} 
  -  f_{out} D  c_{s} \zeta \right)  f_p (1-\cos \theta)^2
  \right] \\
  &-  \dfrac{M_g }{\rho_g  r_{cl}^2 } \left[
  \dfrac{ j  (r_e^2 - r_{cl}^2)}{zFr_{cl}} f_{in} (1-\cos \theta)^2
  \right].\nonumber
\end{align}

Defining variables that are positive constants during each bubble evolution event with constant $j, \, r_e$ and $r_{cl}$,
\begin{equation}
    K_3 = \dfrac{ j  (r_e^2 - r_{cl}^2)}{zFr_{cl}} , \, \, \, \, \, 
    K_4 = \dfrac{f_{out}D c_{s}  }{P_0},
    \label{Aeqn:K3K4}
\end{equation}
the partial derivative of equation~\ref{Aeqn:dthetat3} becomes:
\begin{equation}\label{Aeqn:dthetattheta}
    \dfrac{ \partial }{\partial \theta}  \frac{d\theta}{dt}=\dfrac{M_g }{\rho_g  r_{cl}^2 } \frac{\partial}{\partial \theta}  (1-\cos \theta)^2  \left[
    \left(\dfrac{2 K_4 \gamma \sin \theta}{r_{cl}} - K_4 P_0 \zeta 
    \right)f_p - K_3 f_{in}
    \right].
\end{equation}
At equilibrium, combining equations~\ref{eqn:eq} and \ref{Aeqn:K3K4}, we obtain:
\begin{equation}\label{eqn:eq2}
    \left(2 K_4\gamma  \sin \theta_{eq} - K_4 P_0\zeta r_{cl}
    \right)f_p = K_3 r_{cl} f_{in}.
\end{equation}

\begin{figure}
 \centerline{
 \includegraphics[height=4.5cm]{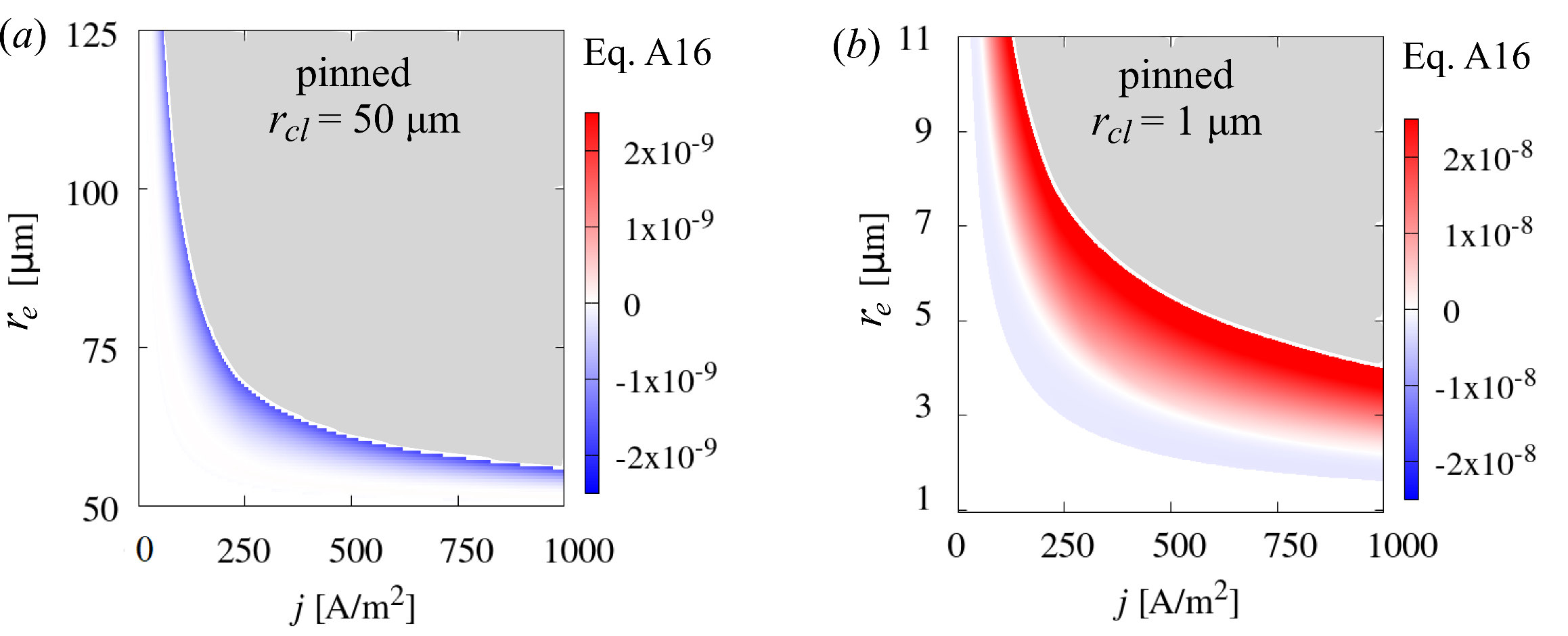}
  }
   \captionsetup{justification=justified, width=\textwidth}
  \caption{
Value of expression (A.16) for two pinned bubbles of different contact radius at dynamic equilibrium.
  The white and grey areas  indicate bubble dissolution and unlimited growth.
  }
\label{fig:dynEqStExpIndicator}
\end{figure}

Combining equation~\ref{eqn:eq2}, equation~\ref{Aeqn:dthetattheta} at equilibrium becomes:
\begin{align}\label{Aeqn:dthetattheta2}
    \left[\dfrac{ \partial }{\partial \theta} \dfrac{d\theta}{dt}\right]_{eq}  &=\dfrac{M_g }{\rho_g  r_{cl}^2 } (1-\cos \theta_{eq})^2  \frac{\partial}{\partial \theta}  \left[
    \left(\dfrac{2 K_4 \gamma \sin \theta}{r_{cl}} - K_4 P_0 \zeta 
    \right)f_p - K_3 f_{in}
    \right]_{eq}\nonumber \\
    &= \dfrac{M_g }{\rho_g  r_{cl}^2 } (1-\cos \theta_{eq})^2 \left[
     \dfrac{2K_4 \gamma}{r_{cl}}f_p\cos\theta + 
     K_3 \left(\dfrac{f_{in}}{f_p}\frac{df_p}{d\theta} - \frac{\partial f_{in}}{ \partial \theta}
     \right)
    \right]_{eq}.
\end{align}
The sign of $ \left[\dfrac{ \partial }{\partial \theta} \dfrac{d\theta}{dt}\right]_{eq} $ is determined by the sign of the square bracket on the right hand side of equation~\ref{Aeqn:dthetattheta2}.
\begin{equation}\label{Aeqn:dthetatthetaS}
     \dfrac{2K_4 \gamma}{r_{cl}}
     f_p (\theta_{eq})\cos\theta_{eq} + 
     K_3 \left[\dfrac{f_{in}}{f_p}\frac{df_p}{d\theta} - \frac{\partial f_{in}}{ \partial \theta}
     \right]_{eq}.
\end{equation}
Unlike the unpinned case, where the sign of $ \left[ \partial (d r_b / dt) / \partial r_b \right]_{eq} $ is clearly negative if $\zeta \leq 0$ (expression~\mbox{\ref{eqn:drtrIndi}}), 
in the pinned case the stability
behavior is more complex.
As shown in figure \ref{fig:dynEqStExpIndicator},
for a relatively large contact radius (50 $\upmu$m), the sign 
of expression \ref{Aeqn:dthetatthetaS}
and thus of \ref{Aeqn:dthetattheta2}
is negative in the parameter range considered.
Thus, larger pinned bubbles at dynamic equilibrium are stable.
For a smaller pinned bubble of 1 $\upmu$m, the sign of expression
\ref{Aeqn:dthetatthetaS}
is found to be positive in the upper equilibrium area where 
$\theta_{eq}>90$\textdegree{}
according to figure \ref{fig:dynEqStPin}, and 
is found to be
negative in the lower equilibrium
area where 
$\theta_{eq}<90$\textdegree{}.
Thus, only in the latter part the dynamic equilibrium is stable.
Considering expression \mbox{\ref{Aeqn:dthetatthetaS}}, we note that its first term is positive and negative when $\theta$ is smaller and larger than 90\textdegree{}, respectively.
Its second term can be assumed to be negative, because $f_{p}$ decreases faster 
with $\theta$ than $f_{in}$ (see figure~\mbox{\ref{fig:finoutp}}).
Therefore, the sign of $ \left[ \partial (d\theta / dt) / \partial \theta \right]_{eq}$ for small bubbles is  determined by the first term of expression \mbox{\ref{Aeqn:dthetatthetaS}} originating from the
Laplace pressure, and for larger
bubbles it is determined by
the second term originating
from the bubble shape.
Note that for bubbles on non-reactive surfaces, $j=0, \, K_3=0$, so that this expression is only determined by the first term.
%
All results of the theoretical solution shown in the figures were
computed by using Matlab.


\section{Variation of surface slip
and diffusion coefficient
}\label{appB}
\begin{figure}
	\centering
	\includegraphics[height=4.5cm]{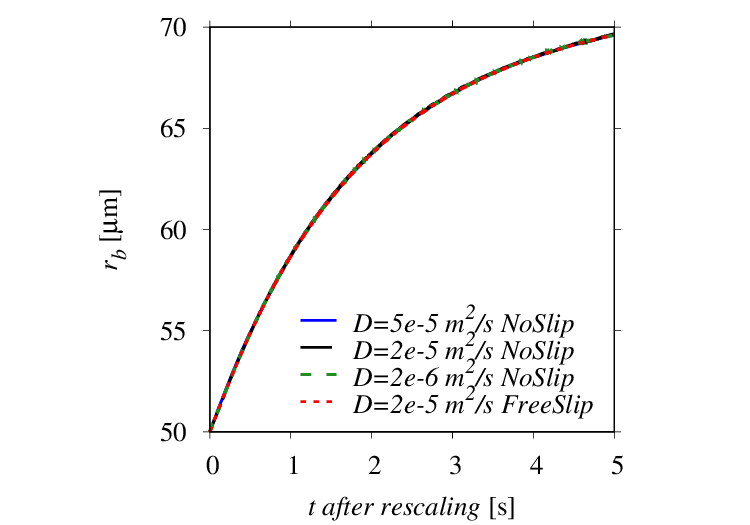}
	 \captionsetup{justification=justified, width=\textwidth}
	\caption{Influence of NoSlip/FreeSlip condition at substrate, and manually enhancing the diffusion coefficient on the evolution of an unpinned bubble.
	The horizontal axis is the time after rescaling according to the ratio of the increased to the real diffusion coefficient.
	$r_{b,ini}=50 \, \upmu$m. $r_{e}= 75 \, \upmu$m, $j=250 $ A/m$^2$, $\theta$=90\textdegree{}.}
\label{fig:FS}		
\end{figure}

As can be seen in figure \ref{fig:FS}, changing the 
boundary condition at the substrate
from no-slip to free-slip does not change the bubble evolution process.
Besides, using different values of manually increased
diffusion coefficients
does not affect the results.
This requires to re-scale time  according to the ratio of increased to real diffusion coefficient.
The enlarged diffusion coefficient
must not impair the dominance
of surface tension
during bubble growth
\citep{han2024}.

\section{Influence of initial radius for unpinned bubbles}\label{sec:Indephy}
\begin{figure}
	\centering
	\captionsetup{justification=justified, width=\textwidth}
	\includegraphics[height=4.5cm]{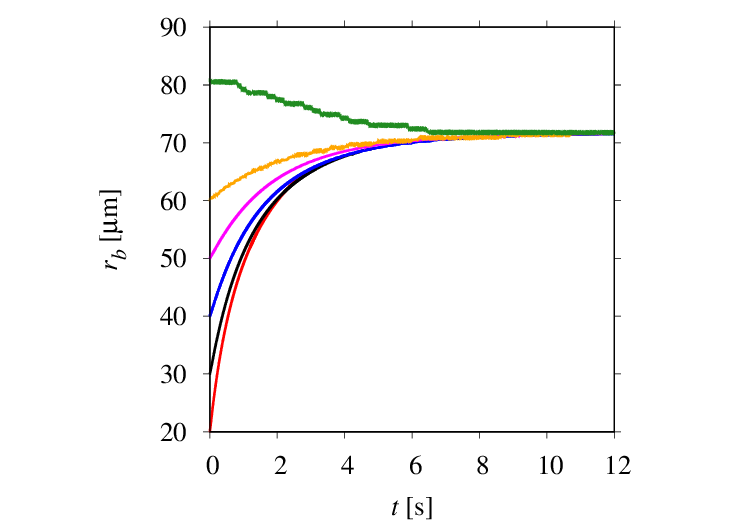} 
	\caption{
	Unpinned bubbles of different initial size
	$r_{b,ini}$, ranging from 20 to 80 $\upmu$m and represented by different colors, 
	evolve towards the same
	dynamic equilibrium state. $j=250 $ A/m$^2$, $r_e=75 \, \upmu$m, $\theta=90$\textdegree{}.
}
\label{fig:rbini}		
\end{figure}

As can be seen in figure \ref{fig:rbini}, unpinned
bubbles of different initial radius
start either to grow or to
dissolve, but eventually are
all approaching the same
state of dynamic equilibrium.
This gives numerical support
to the theoretical stability condition (\ref{eqn:stab0}) 
derived above.
It also indicates that 
in the cases considered here,
convection is not important ($Pe=r_b/D \cdot dr_b/dt \ll 1$), 
as otherwise the final bubble state could be influenced by the different
impact of growth or shrinkage
on the convective mass transfer.

\section{Influence of bulk concentration}\label{sec:Indebulk}
\begin{figure}
 \centerline{
 \includegraphics[height=5cm]{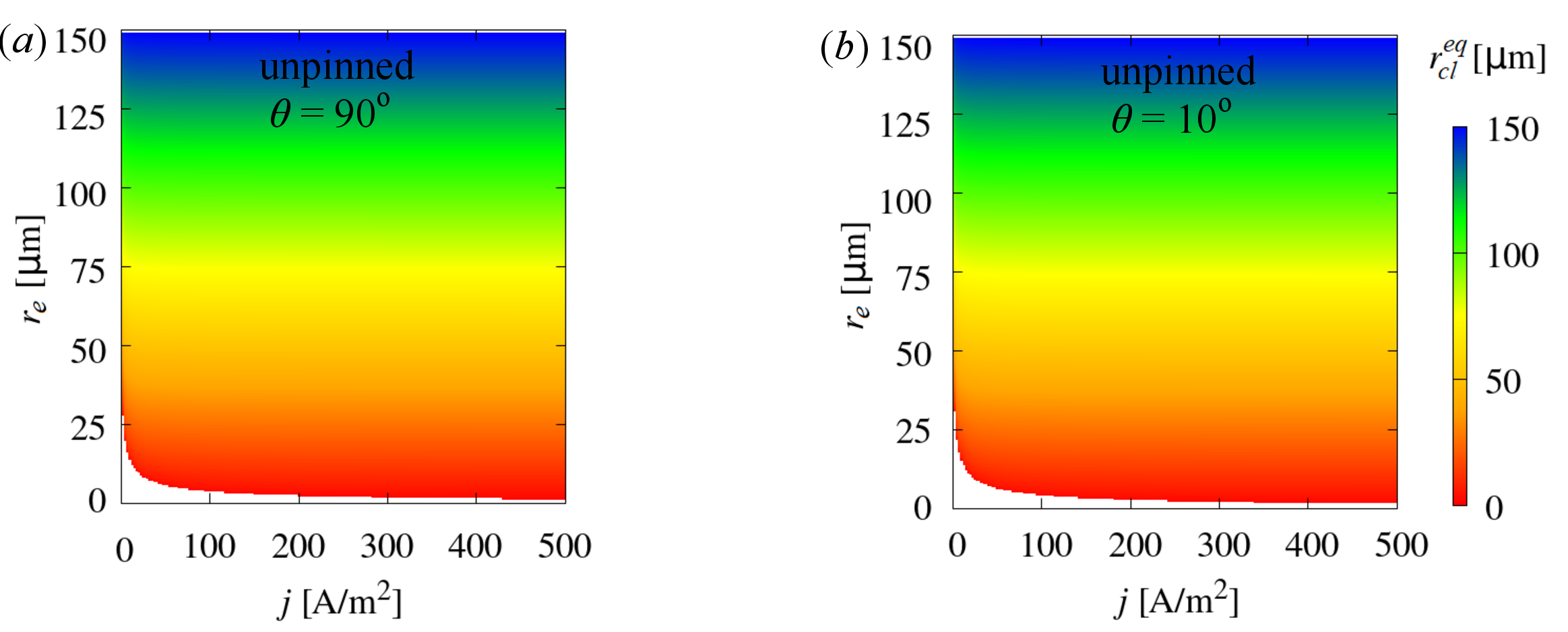}
  }
 \captionsetup{justification=justified, width=\textwidth}
  \caption{Theoretical solution of the equilibrium contact line radius $r_{cl,eq}$ (color surface) for unpinned bubbles 
  versus electrode radius $r_e$ and applied current density $j$ when $c_b=c_s$, i.e.~$\zeta =0$.
  (\textit{a})  $\theta=90$\textdegree{}. (\textit{b})  $\theta=10$\textdegree{}. 
    The bottom white area marks complete dissolution.
  }
\label{fig:dynEqStExpxi0}
\end{figure}

Here we show theoretical results of bubble end states obtained for zero over-saturation, i.e.~$c_b=c_s$ and $\zeta=0$. 
For the unpinned cases, figure~\ref{fig:dynEqStExpxi0}, the current density becomes less important compared to the case of zero bulk concentration ($\zeta=-1$) shown in figure~\ref{fig:dynEqStPin}. 
The equilibrium bubble contact radius approaches the electrode radius in most situations. But for small $j$ and small $r_e$, a dissolution region also appears (white space). 
As in the case of $\zeta=-1$, the critical electrode radius decreases with increasing $j$.

\begin{figure}
 \centerline{
 \includegraphics[height=5cm]{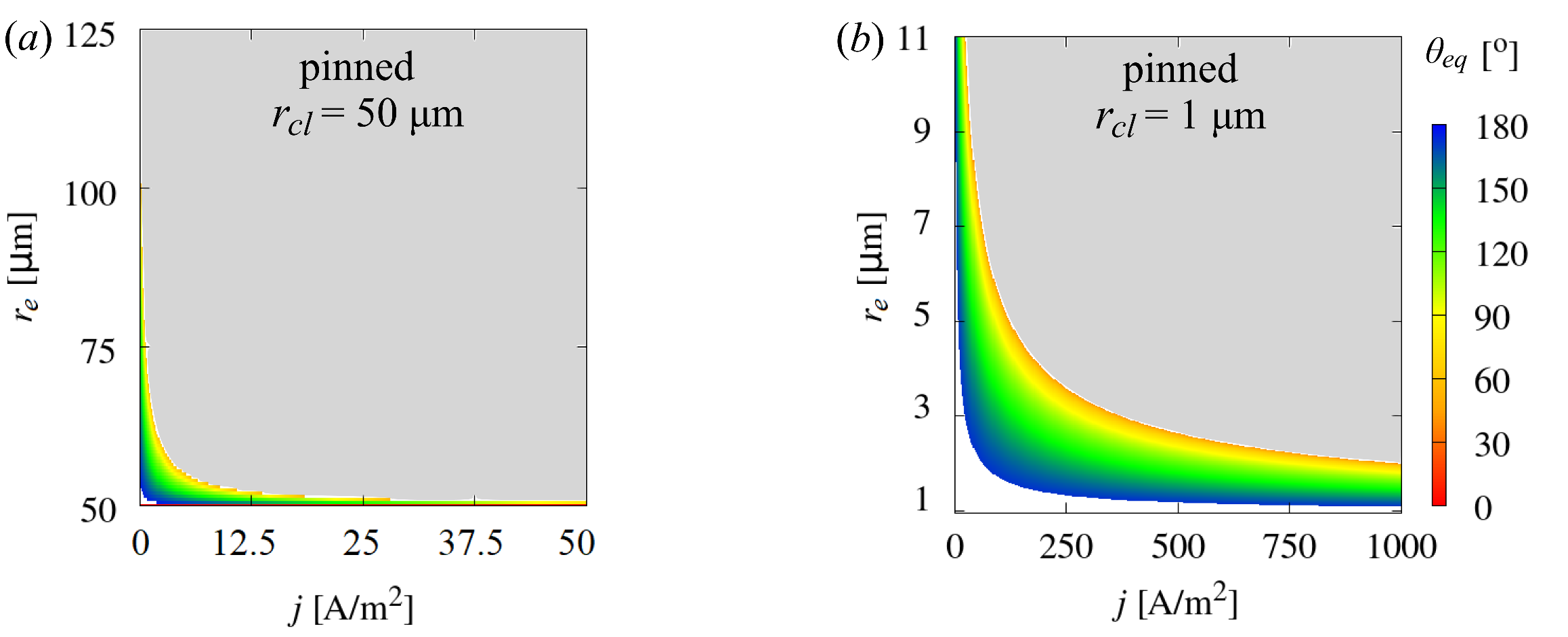}
  }
\captionsetup{justification=justified, width=\textwidth}
  \caption{Theoretical solution of the equilibrium contact angle $\theta_{eq}$ (color surface) 
  for pinned bubbles versus electrode radius $r_e$ and applied current density $j$ when $c_b=c_s$, i.e.~$\zeta =0$.
  (\textit{a}) $r_{cl}=50 \, \upmu$m. 
   (\textit{b}) $r_{cl}= 1 \, \upmu$m.
  The white bottom and grey top areas represent complete dissolution and unlimited growth, respectively.
  }
\label{fig:dynEqStPinxi0}
\end{figure}
For the pinned bubbles, figure~\ref{fig:dynEqStPinxi0}, a dissolution region (white space below the colored surface) and a unlimited-growth-region (grey space above the colored surface) could be observed. The equilibrium contact angle decreases with increasing $r_e$ and $j$. This is qualitatively also similar to the case of $\zeta=-1$ (figure~\ref{fig:dynEqStPin}), but the dissolution and growth regions have become smaller and larger, respectively at the increased bulk concentration.

\bibliographystyle{jfm}
\bibliography{jfm}


\end{document}